\documentstyle[aps,preprint]{revtex}
\newlength{\defaultparindent}
\newenvironment{Default Paragraph Font}{}{}

\begin{document}
\draft
\title{Composite vortex model of the electrodynamics of high-$T_c$ superconductor}
\author{H.-T. S. Lihn and H. D. Drew}
\address{Center for Superconductivity Research, Department of Physics, \\ University
of Maryland, College Park, MD 20742}
\date{\today }
\maketitle

\begin{abstract}
We propose a phenomenological model of vortex dynamics in which the vortex
is taken as a composite object made of two components: the vortex current
which is massless and driven by the Lorentz force, and the vortex core which
is massive and driven by the Magnus force. By combining the characteristics
of the Gittleman-Rosenblum model (Phys. Rev. Lett. {\bf 16}, 734 (1966)) and
Hsu's theory of vortex dynamics (Physica {\bf C 213},305 (1993)), the model
provides a good description of recent far infrared measurements of the
magneto-conductivity tensor of superconducting YBa$_2$Cu$_3$O$_{7-\delta }$
films from 5 cm$^{-1}$ to 200 cm$^{-1}$.\\
\end{abstract}


\narrowtext

With the discovery of high-$T_c$ superconductors and the recognition of
their potential for high field applications much attention has been devoted
to the study of vortex dynamics. Nevertheless, the dissipation mechanism
associated with vortex motion remains poorly understood. From surface
impedance measurements in the MHz and GHz ranges the vortex pinning and
viscosity parameters have been reported based on the Gittleman-Rosenblum
(G-R) model of vortex dynamics \cite{Golosovsky,Pambianchi,Revenaz}. The
observation of magneto-optical activity was reported in far infrared (FIR)
experiments,\cite{Karrai:OA,Choi} which suggests a connection between vortex
dynamics and cyclotron resonance\cite{Drew:Hsu}. This observation greatly
enriches the phenomenology of electrodynamics of vortices. Recently,
quantized vortex core levels in YBa$_2$Cu$_3$O$_{7-\delta }$ (YBCO) have
been probed directly using scanning tunneling microscopy\cite{Maggio}. A
level spacing within the FIR frequency range was reported. Also evidence for
vortex core excitations has been reported in FIR\ experiments\cite{Lihn:gb},
but the role of vortex core structure in vortex dynamics is not yet well
understood. The use of terahertz spectroscopy\cite{Parks,Spielman} and
polarized FIR laser sources\cite{Lihn:KKT} have recently filled in the gap
of measurements between microwave and FIR frequencies. Therefore, a more
satisfactory description of vortex dynamics over wide frequency range is
needed, in order to describe these experimental observations.

Microwave surface impedance studies\cite{Golosovsky,Pambianchi,Revenaz,Parks}
have demonstrated a loss band near zero frequency whose width is
characterized by the depinning frequency $\omega _d$ ($=\kappa /\eta $).
There is remarkable consistency in the reported pinning force constant $%
\kappa $ for YBCO within a factor of two. However, the viscosity $\eta $ was
found to vary much more widely. This leads to a wide range for $\omega _d$.
From measurements of the conductivity up to 800GHz $\omega _d\sim $ 300GHz
(10 cm$^{-1}$) has been reported for YBCO\ films\cite{Parks}. At FIR
frequencies, optical activity has been observed in magneto-transmission
measurements using broadband and laser sources throughout the 5 cm$^{-1}\leq
\omega \leq $ 200 cm$^{-1}$ frequency range\cite{Karrai:OA,Choi,Lihn:KKT}.
The conventional theories based on a Lorentz force acting on the vortices,
e.g. the G-R model\cite{GR} and Coffey-Clem Model\cite{CC}, which are widely
used to describe microwave surface impedance data, can not properly describe
these FIR measurements since they do not produce optical activity\cite{Choi}%
. A model of vortex dynamics proposed by Hsu\cite{Hsu} based on the
microscopic BCS theory in the clean limit gives a Magnus force acting on
vortices and successfully describes the high frequency ($\omega \geq $ 25 cm$%
^{-1}$) optical activity in terms of hybridization of the cyclotron
resonance and the vortex core resonance with the pinning resonance\cite
{Choi,Choi:Vortex}. The hybridized pinning-cyclotron resonance has been
fully resolved in recent FIR laser measurements ($\omega \geq $ 5 cm$^{-1}$)%
\cite{Lihn:KKT}. However, Hsu's theory fails to describe the low frequency
absorption band\cite{Lihn:KKT} observed in the microwave studies\cite
{Golosovsky,Revenaz,Parks}.

In this paper we present a new phenomenological model to describe the
electrodynamics of type II superconductors. The two different inertial
masses that have been discussed in vortex dynamics\cite{Otterlo} suggests a
model consisting two coupled equations of motion which is intended to
capture the qualities of G-R model at low frequencies and Hsu's theory at
higher frequencies. We take the vortex to be a two component composite
object (see Fig.\ref{pinning}). One component is the vortex current outside
the vortex core, which has a G-R type equation of motion, and the other
component is the vortex core with an equation of motion based on Hsu's model%
\cite{Hsu,GR}. The resulting two coupled equations of motion produces a
conductivity tensor with a London term and three Lorentzian oscillators :
one low frequency oscillator similar to the G-R loss band and two finite
frequency chiral oscillators as in Hsu's model. These three resonances are
illustrated schematically in Fig.\ref{poles}. This composite vortex model
provides a good description of the measured magneto-transmission and the
magneto-conductivity tensor from 5 cm$^{-1}\leq \omega \leq $ 200 cm$^{-1}$%
\cite{Lihn:KKT}.

The motion of a vortex line in type-II superconductor under the influence of
an alternating current is limited by frictional and pinning forces. The
simplest description of linear vortex dynamics is \cite{GR} 
\begin{equation}
\vec J\times \widehat{\phi }_0=\eta \,v_v+\kappa \,r_v  \label{GReq}
\end{equation}
where $\vec J=nev_s$ is the driving transport current density ($v_s$ is the
superfluid velocity), $v_v$ $(r_v)$ is the average vortex velocity
(displacement), $\kappa $ is the pinning force constant, $\eta $ is the
viscosity and $\widehat{\phi }_0$ is the flux quantum, $hc/2e$, which is
expressed as a vector in the direction of the applied magnetic field ($%
\widehat{z}$). (A more precise interpretation of $v_v$ will be discussed
later). Eq.(\ref{GReq}) does not contains an inertial term which corresponds
to an assumption of massless dynamics of the vortex current pattern. This
equation of motion produces a pole in the conductivity at zero frequency of
width $\omega _d$ $=\kappa /\eta $ (Peak (2) in Fig.\ref{poles}) in addition
to a London delta function reduced in strength from its zero field value.
This model which is based on classical hydrodynamic analogies is valid only
in those regions outside the core where the superfluid can be described by
the order parameter in the time dependent Ginzberg-Landau\ (TDGL) theory\cite
{Otterlo,Dorsey}. Suhl\cite{Suhl} has considered the inertial mass of the
vortex within the TDGL theory and concluded that it is order $\left( \Delta
/\epsilon _F\right) ^2$ smaller than the naive estimate $M_L\simeq mn\pi \xi
^2$.

A Hall force term $v_v\times \widehat{z}$ can be added to Eq.(\ref{GReq})
leading to 
\begin{equation}
\Lambda \left( v_s-\beta v_v\right) \times \widehat{z}-\eta \,v_v-\kappa
\,r_v=0  \label{GReq1}
\end{equation}
where $\Lambda =nh/2$ the strength of Lorentz force and $\beta $
characterizes the relative strength of the Hall force to the Lorentz force.
When $\beta =1$ the first term corresponds to the Magnus force. A Hall term
is required to describe the anomalous Hall effect observed in both
conventional and high-$T_c$ superconductors in the mixed state where the
sign of dc Hall effect changes if $\beta $ changes its sign\cite
{Dorsey,Hagen}. Otterlo {\it et al.}\cite{Otterlo} have argued that, in a
fermionic system, the strength of the Hall force depends on the electronic
structure through the derivative of the density of states at the Fermi
level. Their theory leads to a small $\beta $ whose magnitude and sign
depends on the details of the electronic band structure at the Fermi level.

From an analysis of vortex motion within the BCS theory Hsu \cite{Hsu}
proposed an equation of motion which has a kinetic term, $\dot v_C$
corresponding to a non-negligeable vortex core mass, a Magnus force $\left(
v_s-v_C\right) \times \widehat{z}$ and a `gauge term' $\dot v_s$ which
couples the vortex motion to the external electric field. Ao and Thouless%
\cite{Ao} have shown that a Magnus force is a general property of vortex
motion from a Berry phase argument. Hsu's equation of motion, normalized by
the vortex core mass, is written as : 
\begin{equation}
\dot v_C=\dot v_s+\Omega \left( v_s-v_C\right) \times \widehat{z}-\frac 1\tau
\,v_C-\alpha ^2\,r_C  \label{Hsueq}
\end{equation}
where $\hbar \Omega $ $\cong \Delta ^2/\epsilon _F$ is the energy spacing of
the quasi particle levels in the vortex core\cite{Spacing}, $v_C$ $(r_C)$ is
the core velocity (displacement), $\frac 1\tau $ is the damping rate
associated with the core motion and $\alpha $ is the bare pinning frequency.
Due to the finite vortex core mass implicit to Eq.(\ref{Hsueq}) two finite
frequency poles of finite width are produced in the conductivity tensor in
addition to a London term. One pole is a hybridized pinning-cyclotron
resonance (Peak (1) in Fig.\ref{poles}) and the other is a hybridized vortex
core-pinning resonance (Peak (3) in Fig.\ref{poles}). However, the vortex
core resonance, which is quenched in the absence of pinning (leaving only
the cyclotron resonance, in accord with Kohn's theorem\cite{Drew:Hsu}),
remains weak for physically reasonable pinning.

The moving vortices produce a Josephson field\cite{Josephson} which couples
to the superfluid by a modification of the London equation : 
\begin{equation}
\dot v_s=\frac{e\vec E}m+\omega _cv_L\times \widehat{z}  \label{Joefield}
\end{equation}
where $\omega _c=eH/mc$ is the cyclotron resonance frequency and $v_L$ is
the average vortex velocity. Elementary arguments show that the Josephson
field due to a moving vortex is proportional to the gradient of the local
magnetic field distribution, $E\sim \nabla B(r)$.\ Since the curvature of
the magnetic field is almost zero in the vortex core\cite{Gygi}, most of the
contribution of the Josephson field should come from the current flow
pattern outside the core, $r\simeq \lambda _L$ (London penetration length),
where there is maximum curvature in the magnetic field distribution of the
vortex. Consequently, we expect $v_L$ to be the velocity of the vortex
current pattern relative to the lattice. Therefore, 
\begin{equation}
v_L=v_v+v_C.  \label{superpose}
\end{equation}
where $v_v$ is interpreted as the velocity of the vortex current pattern
relative to the core and $v_C$ is the velocity of the core relative to the
pinning center, as illustrated in Fig.\ref{pinning}. In this picture the
pinning force $\kappa \,r_v$ acting on the vortex current pattern can be
thought of as arising from the distortion of the vortex through its tendency
to restore itself to the equilibrium configuration. Therefore, in this model 
$\kappa $ is an intrinsic property of the vortex.

As pointed out by Hsu\cite{Hsu}, there is an additional contribution to the
current due to the moving vortex core which acts like a charged particle 
\begin{equation}
\vec J=ne\left( A\,v_s+(1-A)\,v_C\right)  \label{current}
\end{equation}
We determine $A$ by requiring the response function to obey the energy
conservation law. First we neglect the damping terms $\frac 1\tau \,v_C$ and 
$\eta \,v_v$ and make the product $\vec E\cdot \vec J$ energy conserving.
The only nonzero terms are the kinetic energy terms $\frac d{dt}\left( \frac 
12v_s^2\right) $, $\frac d{dt}\left( \frac 12v_C^2\right) $, $\frac d{dt}%
\left( \frac 12v_v^2\right) $, and pinning potential terms $\frac d{dt}%
\left( \frac 12r_C^2\right) $, $\frac d{dt}\left( \frac 12r_v^2\right) $.
Setting all other cross product terms to be zero, we obtain $A=\Omega \left/
\left( \Omega +\omega _c\right) \right. $ which is similar to that obtained
by Hsu\cite{Hsu}. We also find that we have to add a term, $-\omega
_cv_v\times \widehat{z}$, to the right hand side of Eq.(\ref{Hsueq}).
Combined with the `gauge term' $\dot v_s$, $\dot v_s-\omega _cv_v\times 
\widehat{z}$ = $\frac{e\vec E}m+\omega _cv_C\times \widehat{z}$. The
interpretation is that the total electric field acting on the vortex core is
equal to the external electric field plus the Josephson field from the
current flow pattern, but the Josephson field from the core itself is
excluded. However, this is only a small correction since $\omega _c\ll
\Omega $.

From these equations, we can calculate the conductivity function. First from
Eq.(\ref{GReq1}) and Eq.(\ref{Hsueq}) (in left circular polarization) we
define 
\begin{equation}
g_v(\omega ,H)=\frac{v_v}{v_s}=\frac{i\Lambda }{i\beta \Lambda +\eta
+i\kappa /\omega }  \label{vheq}
\end{equation}
and 
\begin{equation}
g_C(\omega ,H)=\frac{v_C}{v_s}=\frac{-i\omega +i\Omega -\omega
_c\,g_v(\omega ,H)}{-i\omega +i\Omega +\frac 1\tau +i\alpha ^2/\omega }
\label{vceq}
\end{equation}
The conductivity function is then 
\begin{equation}
\sigma ^{+}(\omega ,H)=\frac{\vec J}{\vec E}=\frac{ne^2}m\frac{\left[ \Omega
+\omega _c\,g_C(\omega ,H)\right] \left/ \left( \Omega +\omega _c\right)
\right. }{-i\omega -i\omega _c\left[ g_C(\omega ,H)+g_v(\omega ,H)\right] }
\label{conductivity}
\end{equation}
from Eq.(\ref{Joefield}) and Eq.(\ref{current}). $\sigma ^{-}$ is obtained
from $\sigma ^{+}$ through the time reversal symmetry relation $\sigma
^{-}\left( \omega ,H\right) =\sigma ^{+}\left( -\omega ,H\right) ^{*}$\cite
{Lihn:KKT}. This relation suggests a canonical representation of $\sigma $
in which $\sigma ^{+}$ is plotted for positive frequencies and $\sigma ^{-}$
is plotted for negative frequencies.

The conductivity function Eq.(\ref{conductivity}) is causal $(Re[\sigma
]\geq 0)$, satisfies the oscillator strength sum rule, reduces to cyclotron
resonance in the clean limit (in accord with Kohn's theorem\cite{Drew:Hsu})
and further reduces to the London conductivity if the field is set to zero.
With regard to the first two properties, we note that the conductivity
tensor can be resolved into a finite sum of Lorentzian oscillators $\sigma
^{+}(\omega ,H)=$ $ne^2/m\sum_{q=0}^3f_q\left/ \left( -i(\omega -\omega
_q)+\Gamma _q\right) \right. $ but with {\it complex} oscillator strengths, $%
f_q$'s, (except $f_0$ the superfluid oscillator strength is real, see Eq.(%
\ref{eqfs}) below.) and $\sum_{q=0}^3f_q=1$ as in G-R model and Hsu's
theory. However, these {\it complex} oscillator strengths are unusual and
may be an indication that the model needs further refinement. To obtain
cyclotron resonance, we use the Bardeen-Stephen theory\cite{Bardeen} in
which $\eta =B_{c2}\phi _0\sigma _n$ and $\sigma _n=$ $\tau _{qp}c^2/4\pi
\lambda _L^2$ where $\sigma _n$ is the normal state conductivity, and $\tau
_{qp}$ is the quasiparticle scattering rate. In the clean limit where $\tau
_{qp}\rightarrow \infty $ and $\alpha ^2$ = $1/\tau $ = 0, the conductivity
reduces to $-ne^2/m\left/ i(\omega -\omega _c)\right. $.

It is also interesting to consider the normal state limit of the model. In
the flux flow regime where $\alpha ^2$ = 0 but $1/\tau $ $\neq $ $0$, and
near $T_c$ the superfluid carrier is depleted so that $\Lambda ,$ $\Omega
\rightarrow 0$ (or $\ll \eta ,1/\tau $), we obtain $\sigma ^{+}(\omega
,H)\cong $ $ne^2/m\left/ \left( -i(\omega -\omega _c)+1/\tau \right) \right. 
$ which is the form of the metallic cyclotron resonance. However, a
contribution from the quasiparticles should also be included, which would
have the same form but proportional to $n_q\equiv \left( n_{total}-n\right) $%
. Therefore, the model can have a smooth transition into the normal state.

Recently, the FIR magneto-transmission coefficient, $T^{\pm }(\omega ,H)$
for YBCO thin films has been measured over a wide range of frequencies by
the combination of broadband Fourier Transform Spectroscopy (FTS) and CO$_2$
pumped FIR laser measurements\cite{Lihn:KKT}. The magneto-conductivity
tensor was obtained from the magneto-transmission spectra by a
Kramers-Kronig analysis. Fig.\ref{t9} shows the transmission amplitude ratio 
$\left| T^{\pm }(\omega ,H)/T(\omega ,0)\right| ^{1/2}$ at $H$=9T and 4 K
(Panel (a)) and the magneto-conductivity tensor $Re[\sigma ^{+}(\omega ,H)]$
(Panel (b)) in the canonical coordinate representation. A fit to the
composite model (the dotted line) is also shown in Fig.\ref{t9} and is seen
to be in very good agreement with the data except for some small structures
above 50 cm$^{-1}$. These features are induced by the presence of 45$^{\circ
}$ misaligned grains in these samples as discussed in Ref\cite{Lihn:gb}. The
parameters of the best fit are $\alpha =$ 45 cm$^{-1}$, $\frac 1\tau =$ 58 cm%
$^{-1}$, $\Omega =$ 49 cm$^{-1}$, $\omega _c=$ 4.6 cm$^{-1}$ \cite{mass:e},
for the vortex core, and $\beta =$ -0.086, $\kappa =$ 6$\times 10^5$ N/m$^2$%
, $\eta =$ 5.6$\times 10^{-7}$ kg/m sec, for the vortex current, assuming
that $\Lambda =nh/2$ and $ne^2/m=\,c^2/4\pi \lambda _0^2$ where $\lambda _0=$
1850\AA \cite{Lihn:KKT}. By associating the Magnus force constant $M_V\Omega 
$ ($M_V$ is the vortex mass) with $nh/2$ \cite{Ao}, we can also estimate $%
M_V\approx $ $6\times 10^8$ electron mass per cm. The corresponding estimate
for the coherence length $\xi _V\cong $ 26\AA\ by taking $M_V=mn\pi \xi _V^2$%
\cite{size}. The fitting also gives $\omega _d=\kappa /\eta \approx $ 10 cm$%
^{-1}$, very close to the value reported by Parks {\it et al.}\cite{Parks}.

The resulting conductivity function is seen to be dominated by two major
oscillators in addition to the London term ($\omega _0=0$, $\Gamma _0=0$)%
\cite{Lihn:KKT}: a low frequency oscillator at $\sim $3 cm$^{-1}$ and width
of 10 cm$^{-1}$, associated with the vortex currents; and a high frequency
chiral oscillator at -24 cm$^{-1}$ and width of 17 cm$^{-1}$ in the hCP mode
which is associated with the vortex core (the hybridized pinning-cyclotron
resonance). Another oscillator has been identified at about 65 cm$^{-1}$ in
the eCP mode (the hybridized vortex core-pinning resonance), but it is very
weak because of motional quenching\cite{Lihn:gb}.

The best fit also gives a negative $\beta $ which may be related to the sign
reversal of the dc Hall effect\cite{Otterlo,Dorsey,Hagen}. The presence of
the Hall force ($\beta \neq 0$) in Eq.(\ref{GReq1}) shifts the low frequency
oscillator to finite frequency $\omega _1\approx -\beta \Lambda \,\Gamma
_1/\eta $ (neglecting the contributions related to the vortex core, which is
a fairly good approximation in this case.), where $\Gamma _1\approx (\kappa
+\Lambda \omega _c)/\eta $ is the width of this oscillator. $\beta <0$
results from $\omega _1>0$, which means that the low frequency oscillator is
slightly electron-like (eCP). In the experimental paper\cite{Lihn:KKT} we
also pointed out that the superconvergent sum rule, $\int_0^\infty \ln
\left| T^{+}(\omega ,H)/T^{-}(\omega ,H)\right| \frac{d\,\omega }\omega =0$,
which is a consequence of the supercurrent, is not satisfied unless the low
frequency oscillator is shifted toward positive frequencies (eCP) to avoid
the excessive low frequency weight in the optical activity.

Finally we discuss $f_0$ the oscillator strength of the superfluid
condensate within the composite vortex model. In the low frequency limit $%
\lim_{\omega \rightarrow 0}\sigma ^{+}(\omega ,H)=$ $\frac{ne^2}m%
f_0(H)\left/ \left( -i\omega \right) \right. $, where 
\begin{equation}
f_0(H)=\frac{\Omega /(\Omega +\omega _c)}{1+\omega _c\left( \Omega /\alpha
^2+\Lambda /\kappa \right) }.  \label{eqfs}
\end{equation}
Note that $f_0(H)=$ $\lambda _L^2/\lambda ^2(H)$ which can be measured in
microwave experiments and that $f_0$ is independent of the damping
parameters $\eta $ and $1/\tau $. In the G-R model, $f_0(H)=$ $\left(
1+nh\omega _c/2\kappa _{GR}\right) ^{-1}$. Comparing Eq.(\ref{eqfs}) with
the G-R result we can define an effective $``\kappa _{GR}"$ in the low field
limit where $f_0(H)\cong 1-\omega _c(\Omega /\alpha ^2+\Lambda /\kappa
+1/\Omega )$. Taking $M_V\Omega $ $=\Lambda $ $=nh/2$ we find $1/``\kappa
_{GR}"=1/\kappa +1/\kappa _C+1/\Lambda \Omega $ where $\kappa $ is the force
constant associated with the force between the vortex currents and the core
and $\kappa _C=M_V\alpha ^2$ is the force constant associated with the force
between the core and the pinning center. In our model $\kappa $ is an
intrinsic property of the vortex. A review of the published experiment data%
\cite{Golosovsky,Pambianchi,Revenaz,Parks} indicates that the observed $%
\kappa $ is universal within a factor of two, which is difficult to
understand in terms of pinning to random defects in the lattice. The last
term $1/\Lambda \Omega $ is also intrinsic and comes from the reduction in
the condensate density due to the vortex core volume ($\kappa /\Lambda
\Omega \equiv 4\pi \kappa m\xi _V^2/nh^2\approx 0.12$). In general, we see
that the small ``force constant'' dominates $f_0$. Therefore for strong
pinning, such as occurs in YBCO films, the intrinsic pinning may determine $%
``\kappa _{GR}"\cong \kappa $ (in our case, $\kappa /\kappa _C\approx 0.13$%
), which may explain the observed universal value reported in the
literature. As $\kappa _C$ gets small, for example in high quality single
crystals, the extrinsic pinning $\kappa _C$ may dominate and $``\kappa
_{GR}" $ would become smaller and sample dependent.

The authors acknowledged helpful discussions with T. C. Hsu, S. G. Kaplan,
J. Orenstein, S. Wu, and V. Yakovenko. This work was supported in part by
National Science Foundation under Grant No. DMR 9223217.

\begin{figure}[tbp]
\caption{The vortex core (grey circle in the center) is pinned to the
lattice defect (cross). The circulating vortex current pattern (concentric
annular rings) is bound to the vortex core by the intrinsic restoring force
(zigzag springs).}
\label{pinning}
\end{figure}

\begin{figure}[tbp]
\caption{Three poles in the conductivity of the composite vortex model. Peak
(1) and (3) come from the magneto-response of the vortex core. Peak (2)
comes from the massless vortex current flow around the vortex core. The peak
positions and oscillator strengths are the same as in Fig.\ref{t9}(b) but
made much sharper for illustration purposes. hCP (eCP) stands for hole
(electron) cyclotron resonance polarization mode.}
\label{poles}
\end{figure}

\begin{figure}[tbp]
\caption{(a) The transmission amplitude ratio $\left| T^{\pm }(\omega
,H)/T(\omega ,0)\right| ^{1/2}$ as a function of frequency $\omega $ for
YBCO thin films at $H$=9T and 4 K. (b) The magneto-conductivity $Re[\sigma
^{+}(\omega ,H)]$ as a function of frequency $\omega $. The solid line
represents the broadband data from 30 cm$^{-1}$ to 200 cm$^{-1}$ and the
triangular points represent data from the laser source from 24.5 cm$^{-1}$
down to 5.26 cm$^{-1}$. The dotted line is the fit to the composite vortex
model. The dashed line in (b) is the residual metallic background $Re[\sigma
(\omega ,0)]$.}
\label{t9}
\end{figure}


\begin{references}
\bibitem{Golosovsky}  M. Golosovsky, M. Tsindlekht, H. Chayet, and D.
Davidov, Phys. Rev. {\bf B} {\bf 50}, 470 (1994).

\bibitem{Pambianchi}  M. S. Pambianchi, D. H. Wu, L. Ganapathi, and S. M.
Anlage, IEEE {\bf 3}, 2774 (1992).

\bibitem{Revenaz}  S. Revenaz {\it et al.}, Phys. Rev. {\bf B 50}, 1178
(1994).

\bibitem{Karrai:OA}  K. Karrai {\it et al.}, Phys. Rev. Lett. {\bf 69}, 355
(1992).

\bibitem{Choi}  E.-J. Choi, H.-T. S. Lihn, H. D. Drew, and T. C. Hsu, Phys.
Rev. {\bf B 49}, 13271 (1994).

\bibitem{Drew:Hsu}  H. D. Drew and T. C. Hsu, Phys. Rev. {\bf B 52}, 9178
(1995).

\bibitem{Maggio}  I. Maggio-Aprile {\it et al.}, Phys. Rev. Lett. {\bf 75},
2754 (1995).

\bibitem{Lihn:gb}  H.-T. S. Lihn {\it et al.}, Phys. Rev. {\bf B 53}, 1
(1996).

\bibitem{Parks}  B. Parks {\it et al.}, Phys. Rev. Lett {\bf 74},3265 (1995).

\bibitem{Spielman}  S. Spielman {\it et al.}, Phys. Rev. Lett. {\bf 73},
1537 (1994).

\bibitem{Lihn:KKT}  H.-T. S. Lihn {\it et al.}, submitted for publication.
Preprint Library, cond-mat (9510045).

\bibitem{GR}  J. I. Gittleman and B. Rosenblum, Phys. Rev. Lett. {\bf 16},
734 (1966).

\bibitem{CC}  M. W. Coffey and J. R. Clem, Phys. Rev. Lett. {\bf 67}, 386
(1991).

\bibitem{Hsu}  T. C. Hsu, Physica {\bf C 213},305 (1993).

\bibitem{Choi:Vortex}  E.-J. Choi {\it et al.}, Physica {\bf C 254}, 258
(1995).

\bibitem{Otterlo}  A. van Otterlo {\it et al.}, Phys. Rev. Lett {\bf 75}%
,3736 (1995).

\bibitem{Dorsey}  A. T. Dorsey, Phys. Rev. {\bf B 46}, 8376 (1992).

\bibitem{Suhl}  H. Suhl, Phys. Rev. Lett. {\bf 14}, 226 (1965).

\bibitem{Hagen}  S. J. Hagen {\it et al.}, Phys. Rev. {\bf B 47}, 1064
(1993).

\bibitem{Ao}  Ping Ao and D. J. Thouless, Phys. Rev. Lett. {\bf 70}, 2158
(1993).

\bibitem{Spacing}  In order to make an energy-conserving theory, the $\Omega 
$ used in this paper corresponds to $\bar \Omega _0\equiv \Omega _0(1-\Phi )$
in Ref.\cite{Choi}.

\bibitem{Josephson}  B. D. Josephson, Phys. Lett. {\bf 16},242 (1965).

\bibitem{Gygi}  F. Gygi and M. Schluter, Phys. Rev {\bf B 43}, 7609 (1991).

\bibitem{Bardeen}  J. Bardeen and M. J. Stephen, Phys. Rev. {\bf 140}, A1197
(1965).

\bibitem{mass:e}  This $\omega _c$ corresponds to the effective mass $m^{*}$
of 1.8 $m_e$. In high frequency measurements\cite{Choi}, we have deduced a
mass of 3.1 $m_e$. This discrepency, which is outside of experimental error,
is related to $\beta \neq 0$ in the composite vortex model, which also gives
some high frequency optical activity. The interpretation of this is not yet
clear.

\bibitem{size}  $\pi \xi _V^2$ assumes the cylinder approximation for the
vortex core.
\end{references}
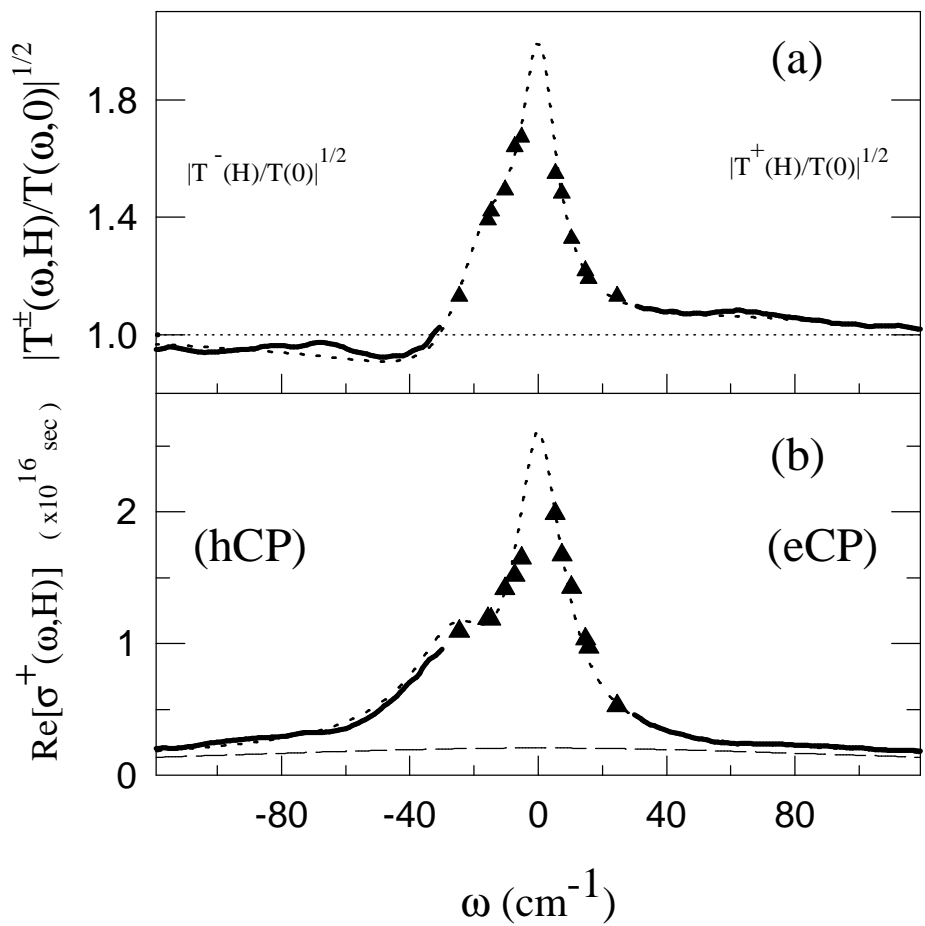
\end{document}